\documentclass[aps,showpacs,12pt,notitlepage]{revtex4-1}
\usepackage{amsmath}
\usepackage{latexsym}
\def\[{\left\lbrack}
\def\]{\right\rbrack}

\def\({\left(}
\def\){\right)}
\newcommand{\be}{\begin{equation}}
\newcommand{\ee}{\end{equation}}
\newcommand{\ea}{\end{eqnarray}}
\newcommand{\ba}{\begin{eqnarray}}

\begin{document}

\title{Hidden symmetries in the two-dimensional isotropic antiferromagnet}

\author{S. A. Leonel$^a$\footnote{\sf E-mail: sidiney@fisica.ufjf.br}, A. C. R. Mendes$^a$\footnote{\sf E-mail: albert@fisica.ufjf.br}, W. Oliveira$^a$\footnote{\sf E-mail: wilson@fisica.ufjf.br}, G. L. Silva$^a$\footnote{\sf E-mail: glima@fisica.ufjf.br}, and L. M. V. Xavier$^a$\footnote{\sf E-mail: lucianaxavier@fisica.ufjf.br}}
\affiliation{${}^{a}$Departamento de F\'{\i}sica,ICE, Universidade Federal de Juiz de Fora,\\
36036-330, Juiz de Fora, MG, Brazil\\
\today\\}
\pacs{11.15.-q; 11.10.Ef; 75.10.Hk; 75.10.Jm}

\begin{abstract}

  We discuss the two-dimensional isotropic antiferromagnet in the
  framework of gauge invariance. Gauge invariance is one of the most
  subtle useful concepts in theoretical physics, since it allows one
  to describe the time evolution of complex physical systesm in
  arbitrary sequences of reference frames. All theories of the
  fundamental interactions rely on gauge invariance. In Dirac's
  approach, the two-dimensional isotropic antiferromagnet is subject
  to second class constraints, which are independent of the
  Hamiltonian symmetries and can be used to eliminate certain
  canonical variables from the theory. We have used the symplectic
  embedding formalism developed by a few of us to make the system
  under study gauge-invariant. After carrying out the embedding and
  Dirac analysis, we systematically show how second class constraints
  can generate hidden symmetries. We obtain the invariant second-order
  Lagrangian and the gauge-invariant model Hamiltonian. Finally, for a
  particular choice of factor ordering, we derive the functional
  Schr\"odinger equations for the original Hamiltonian and for the
  first class Hamiltonian and show them to be identical, which
  justifies our choice of factor ordering.
\end{abstract}

\maketitle

\section{Introduction}

Since the discovery of high-$T_c$ superconductivity \cite{Mag1},
interest in the two-dimensional Heisenberg magnets has been
intensively revived. The study of magnetism in two dimensions (2D) has motivated much
theoretical and experimental work \cite{Mag2, Mag3, Mag4} and led to
substantial progress in the understanding of 2D magnetism \cite{Mag5,
  Mag6, Mag7}. This includes the $O(3)$ non-linear sigma model, which
describes the continuum classical limit of the 2D
Heisenberg antiferromagnet.

A consistent, systematic study of constrained systems was first
established by Dirac \cite{Dirac}. The main goal of the so-called
Dirac formalism was to obtain the Dirac brackets, the bridge to the
commutators in quantum theory. With its categorization of constraints
as of first or second class, primary or secondary, etc., this
formalism has become one of the standards for the analysis of
constrained theories. Faddeev and Jackiw \cite{JackiwA} proposed a
first-order Lagrangian geometric method for the symplectic
quantization of constrained systems, which is different from the
traditional Hamiltonian Dirac approach. In the Faddeev-Jackiw method
we need not introduce primary constraints as in the Dirac formalism,
which stems from the definition of the canonical momenta. Nor is it
necessary to classify constraints as of first or second class, primary or
secondary. All the constraints are held to
the same standard. Barcelos-Neto and Wotzasek proposed the
symplectic formalism \cite{Barcelos}, an improved version of the
Faddeev-Jackiw method for the case in which the constraints are only
partially eliminated.

The central goal of the symplectic formalism is to turn the system
into a first-order Lagrangian with certain auxiliary fields, the
definition of that Lagrangian being independent of how the auxiliary
fields are introduced. The first-order Lagrangian, which consists of a
few symplectic variables and their generalized canonical momenta,
gives the symplectic two-form matrix $f_{ij}$. In the symplectic
formalism the system can be classified as constrained or
unconstrained, depending on the singular behavior of the symplectic
two-form matrix. As a result, the algorithm runs into one of three
alternatives. In the first alternative, if the symplectic two-form
matrix is nonsingular, it can be inverted to yield generalized
brackets, which correspond to Dirac brackets.

In the second alternative, if the symplectic two-form matrix is singular, there
may be some non-trivial zero-mode, which generates the constraints in
the context of the symplectic quantization method. The constraints can
then be transported to the canonical sector by means of appropriate
Lagrangian multipliers and are regarded as conjugate canonical
one-form components, whereas the Lagrange multipliers are treated as
symplectic variables. With this new first-order Lagrangian, a finite
number of iterations usually suffices to make the symplectic two-form
matrix non-singular and to yield the generalized brackets of the
symplectic variables, which coincide with those in the Dirac
formalism.  

Finally, even with a singular two-form matrix, it may be that the
original zero-modes impose no constraints on the dynamical variables
at the first stage of iteration. In the absence of additional
constraints, the original canonical sector in the first-order
Lagrangian is unchanged, and we can say that the system has a gauge
symmetry with the field-transformation rules supported by the
zero-modes.

Recently, the functional Schr\"odinger representation has been
systematically used to quantize different field
theories. Different theories have derived divers predictions, part of which
are physically appealing, from the wave-functionals
obtained so far. The so-called \emph{vacuum angle} is one of
the important theoretical features of gauge theories obtained from the functional
Schr\"odinger representation with no instanton approximation \cite{Jackiw1}.

We will make use of a general canonical formalism of embedding
developed by a few of us on the basis of the symplectic formalism \cite{Oliveira},
which embeds a second-class system into one with gauge
invariance. The first-class Hamiltonian leads to the same classical
theory as the original one. We shall then derive the functional
Schr\"odinger equation using the Dirac first-class quantization
formalism \cite{Dirac}.

This paper is organized as follows. In Section \ref{s2}, we obtain the
Dirac brackets of the model, via two distinct methods: the usual Dirac
algorithm and the symplectic formalism. In Section \ref{s3}, we embed
the isotropic antiferromagnet into a gauge-symmetric system and
discuss the interesting physical situation in which a second class
constraint acts as symmetry generator of the hidden symmetries of the
model. In Section \ref{s4}, we derive the functional Schr\"odinger
equations of the original and first-class models and show that they
are all identical. In Section \ref{s5}, we present our concluding
remaks and future perspectives. Finally, in Appendices~\ref{A1}~and
\ref{A2}, we briefly review the symplectic formalism and present the
general theory of symplectic embedding, respectively. For briefness,
throughout the paper we write ``Lagrangian'' and ``Hamiltonian'' for the
Lagrangian and Hamiltonian densities, respectively.

\section{Dirac brackets}\label{s2}

\subsection{Dirac brackets via the Dirac formalism}

One of the simplest theoretical descriptions of a 2D
isotropic antiferromagnet at low temperatures is given by the
Lagrangian
\begin{equation}
{\cal L'}=\frac{J}{2}\,\partial_\mu{S}^i\partial^\mu{S}_i\qquad(\mu = 0,1,2),
\label{BJP1}
\end{equation}
where $\vec{S} = (S_1, S_2, S_3)$ is a three dimensional vector
subjected to the constraint $S_i^2 = 1$ and satisfies the usual spin
Poisson bracket. The metric has signature $(+,-,-)$ and
we are using the convention of sum over repeated indices. After a Legendre
transformation, the canonical Hamiltonian is given by the expression
\begin{equation}
\label{BJP2}
{\cal H'} = \frac{\pi_{i}^2}{2J} + \frac{J}{2}\,(\partial_k S_i)^2,
\end{equation}
where $k = 1,2$ and $\pi_i$ is the momenta canonically conjugate to
the coordinates $S_i$. 

We are particularly interested in the fields satisfying a condition that
follows from the fact that the energy is finite (static solution). In
this case, the Hamiltonian that follows from the Lagrangian
(\ref{BJP1}) is $ {\cal H''} = \frac{J}{2}\,(\partial_k
S_i)^2, $ which is the classical continuous version, valid at low
temperatures, of the 2D isotropic antiferromagnetic Heisenberg model
\cite{Polyakov}, described by the quantum discrete Hamiltonian
\begin{equation}
\label{BJP3}
{\it H} = -J \sum_{<i,j>} \vec{S_i} \cdot \vec{S_j}, \qquad(J < 0),
\end{equation}
where $i,j$ represent lattice cites and the bracket in the sum
indicates nearest neighbors, and the fields satisfy the constraint
$S_i^2 = 1$. Note that the first term on the right-hand side of Eq. (\ref{BJP2}) can be dropped since Lorentz invariance allows the moving finite-$\pi$ solution to be obtained by a boost of the static $\pi = 0$ solution. For details, the interested reader can consult \cite{Haldane}. ${\cal H''}$ is also the Hamiltonian for the $O(3)$ non-linear sigma model.

Instead of Eq.~(\ref{BJP1}), consider the Lagrangian 
\begin{equation}
{\cal L}=\frac{J}{2}\,\partial_\mu{S}^i\partial^\mu{S}_i
- \frac{1}{2}\,\lambda\,\bigl({S}^i{S}_i - 1\bigr),
\label{3001}
\end{equation}
where we have introduced a Lagrange multiplier $\lambda$ to account
for the constraint $S_i^2 = 1$. 

From (\ref{3001}) we obtain the
Hamiltonian
\begin{equation}
\label{3001a}
{\cal H} = \frac{\pi_{i}^2}{2J} + \frac{J}{2}\,(\partial_k S_i)^2 + \frac{1}{2}\,\lambda\,\bigl({S}^2_i - 1\bigr).
\end{equation}

From the persistence in time of the constraint $\Omega_1 = {S}^2_i -
1$, we obtain the constraint $\Omega_2 = S_i \pi^i$. In the
Dirac-Hamiltonian approach these second-class constraints are the only
constraints on the system. The nonsingular constraint matrix $C$ is
given by the equality
\begin{equation}
\label{3001b}
C = 
\begin{pmatrix}
0  & 2 \\
-2 & 0 \\
\end{pmatrix}
\,{S}^2_i \delta(x - y).
\end{equation}

The inverse of $C$ yields the usual Dirac brackets of the theory, namely
\begin{align}
\label{3001c}
\lbrace S_i(x),S_j(y)\rbrace^* &= 0, \nonumber\\
\lbrace S_i(x),\pi_j(y)\rbrace^* &= \(\delta_{ij} - \frac{S_{i}S_{j}}{{S}^2_i}\)\delta(x - y),\\
\lbrace \pi_i(x),\pi_j(y)\rbrace^* &= \frac{(S_i\pi_j - S_j\pi_i)}{{S}^2_i}\delta(x - y).\nonumber
\end{align}

\subsection{Dirac brackets via the symplectic formalism}

To implement the symplectic method let us introduce the auxiliary
variables $\pi^i$, so that the original second-order Lagrangian in the
velocity, Eq.~(\ref{3001}), can be written as the first-order Lagrangian 
\begin{equation}
\label{3002}
{\cal L}^{(0)} = \pi^i\dot{S}_i - V^{(0)},
\end{equation}
with
\be
\label{3002a}
V^{(0)} = \frac{1}{2J}\,\pi^2_i + \frac{1}{2}\,\lambda\,\bigl({S}^2_i - 1\bigr) - \frac{J}{2}\,(\partial_k S_i)^2.
\ee

The symplectic coordinates are $\xi_\alpha^{(0)}=(S_i,\pi_i,\lambda)$
and the superscript ${(0)}$ indicates that we are at iteration zero. The symplectic tensor given by
Eq.~(\ref{2010}) is computed in this case as
\begin{equation}
\label{3003}
f^{(0)} = \left(
\begin{array}{ccc}
0           & -\delta^{ij} & 0 \\
\delta^{ji} &         0     & 0 \\
0           &         0     & 0
\end{array}
\right)\,\delta(x - y).
\end{equation}

This matrix being singular, we consider the following zero-mode,
\begin{equation}
\label{3004}
\nu^{(0)} = \left(
\begin{array}{ccc}
{\bf 0} \\
{\bf 0} \\
1
\end{array}
\right).
\end{equation}
Contraction of this zero-mode with the gradient of the symplectic
potential $V^{(0)}$ in Eq.~(\ref{3002a}) yields the following constraint
\be
\label{3005a}
\Omega_1 = \frac{1}{2}\,(S^2_i- 1).
\ee

To follow the symplectic formalism we must introduce this constraint
into the canonical sector of the first-order Lagrangian~(\ref{3002})
by means of a Lagrange multiplier $\rho$, which yields the
first-iteration Lagrangian
\begin{equation}
\label{3005}
{\cal L}^{(1)} = \pi^i\dot{S}_i + \Omega_1\dot{\rho} - V^{(1)}
\end{equation}
with
\begin{equation}
\label{3006}
 V^{(1)}=V^{0}\mid_{\Omega_1=0} =\frac{1}{2J}\,\pi^2_i  + \frac{J}{2}\,(\partial_k S_i)^2.
\end{equation}

The new symplectic coordinates are $\xi_\alpha^{(1)}=(Si_i,\pi_i,\rho)$, with the following one-form canonical momenta,
\begin{align}
\label{3007}
A_{S_i}^{(1)} &= \pi^i,\quad A_{\pi_i}^{(1)} = 0,\quad A_{\rho}^{(1)} = \frac{1}{2}\,(S^2_i- 1).
\end{align}

The corresponding symplectic tensor $f^{(1)}$ is given by the matrix
\begin{equation}
\label{3008}
f^{(1)}=
\begin{pmatrix}
0           & -\delta^{ij} & S^i \\
\delta^{ji} &         0     & 0 \\
-S^j          &         0     & 0
\end{pmatrix}
\,\delta(x - y),
\end{equation}
which is singular and has a zero-mode that generates a new constraint, 
\begin{equation}
\label{3009}
\Omega_2 = \frac{1}{J}\,S_k\pi^k.
\end{equation}

We next introduce the constraint $\Omega_2$ into the first-iteration
Lagrangian~(\ref{3005}) with a Lagrange multiplier $\zeta$, to obtain the
second-iteration Lagrangian
\begin{equation}
\label{3010}
{\cal L}^{(2)} = \pi^i \dot{S}_i + \frac{1}{2}(S^2_i- 1)\dot{\rho} + \frac{1}{J} (S_i\pi^i) \dot{\zeta}- V^{(2)},
\end{equation}
with $V^{(2)}$ = $V^{(1)}\mid_{\Omega_1=0}$. The symplectic
coordinates are now $\xi_\alpha^{(2)}=(S_i,\pi_i,\rho,\zeta)$, and the
new one-form canonical momenta are
\begin{align}
\label{formula22}
A_{S_i}^{(2)} &= \pi_i,&\,A_{\pi_i}^{(2)} = 0, \nonumber \\
A_{\rho}^{(2)} &= \frac{1}{2}\,S^2_i - 1,&\,A_{\zeta}^{(2)} = \frac{1}{J}\,S_i\pi^i.
\end{align}

The corresponding matrix $f^{(2)}$ is
\begin{equation}
\label{3011}
f^{(2)}=\left(
\begin{array}{cccc}
0           & -\delta^{ij}  &  S^i   &   \frac{1}{J}\pi^i \\
\delta^{ji} &         0     &   0    &    \frac{1}{J}S^i \\
-S^j         &         0     &   0    &     0   \\
-\frac{1}{J}\pi^j       &        -\frac{1}{J}S^j    &   0    &     0
\end{array}
\right)\,\delta(x - y),
\end{equation}
which is nonsingular. We immediately identify the Dirac brackets,
given by the equalities
\begin{align}
\lbrace S_i(x),S_j(y)\rbrace^* =&\lbrace \rho(x),\rho(y)\rbrace^* = \lbrace \zeta(x),\zeta(y)\rbrace^* = 0, \nonumber\\
\lbrace S_i(x),\pi_j(y)\rbrace^* &= \(\delta_{ij} - \frac{S_{i}S_{j}}{{S}^2_i}\)\delta(x - y),\nonumber\\
\lbrace \pi_i(x),\pi_j(y)\rbrace^* &= \frac{(S_i\pi_j - S_j\pi_i)}{{S}^2_i}\delta(x - y).\\
\lbrace S_i(x),\rho(y)\rbrace^* &= - \frac{S_i}{{S}^2_i}\delta(x - y),\nonumber\\
\lbrace \pi_i(x),\rho(y)\rbrace^* &= \frac{\pi_i}{{S}^2_i}\delta(x - y),\nonumber\\
\lbrace \pi_i(x),\zeta(y)\rbrace^* &= - \frac{J S_i}{{S}^2_i}\delta(x - y),\nonumber
\end{align}
which coincide with the results in Ref.~(\ref{3001c}) for the variables $S_i$ and $\pi_i$.
In principle this indicates that the model lacks gauge simmetry.

\section{Hidden symmetries in the 2D isotropic antiferromagnet}
\label{s3}

To disclose the hidden symmetry in the $2D$ isotropic antiferromagnet,
again following the symplectic embedding formalism, 
we now extend the original phase space with a Wess-Zumino (WZ)
field. To this end we introduce two arbitrary functions $\Psi(S_i,\pi_i,\theta)$ and  $G(S_i,\pi_i,\theta)$ into the
first-order Lagrangian as follows:
\begin{equation}
\label{3012a}
{\tilde {\cal L}}^{(0)} = \pi^i\dot{S}_i + \Psi\dot{\theta} - {\tilde V}^{(0)},
\end{equation}
where the symplectic potential is

\begin{equation}
\label{3013a}
{\tilde V}^{(0)} = \frac{1}{2J}\,\pi^2_i + \frac{1}{2}\,\lambda\,\bigl(S^2_i - 1\bigr) + \frac{J}{2}\,(\partial_k S_i)^2 + G(S_i,\pi_i,\theta),
\end{equation}
with $G(S_i,\pi_i,\theta)$ satisfying
Eqs.~(\ref{2060})~and~(\ref{2070}), in Appendix~\ref{A2}.

The symplectic coordinates are $\tilde \xi_\alpha^{(0)}=(S_i,\pi_i,\lambda,\theta)$,
with the following one-form canonical momenta:
\begin{align}
\label{3015}
\tilde A_{S_i}^{(0)} &= \pi_i,\quad\tilde A_{\pi_i}^{(0)} = 0, \nonumber \\
\tilde A_{\lambda}^{(0)} &= 0,\quad\tilde A_{\theta}^{(0)} = \Psi.
\end{align}

As dictated by the symplectic embedding formalism, the corresponding
matrix $\tilde f^{(0)}$, given by the equality
\begin{equation}
\label{3016}
\tilde f^{(0)} =\begin{pmatrix}
0  & -\delta^{ij}  &  0  &   \dfrac{\partial\Psi_y}{\partial{S}^{x}_i} \\
\delta_{ji} &  0 & 0  & \dfrac{\partial\Psi_y}{\partial\pi^{x}_i} \\
0 & 0 & 0  & \dfrac{\partial\Psi_y}{\partial\lambda^{x}} \\
- \dfrac{\partial\Psi_x}{\partial{S}^{y}_j} & -\dfrac{\partial\Psi_x}{\partial\pi^{y}_j}  & -\dfrac{\partial\Psi_x}{\partial\lambda^{y}} & 0
\end{pmatrix}\,
\delta(x-y),
\end{equation}
must be singular, which leads to
${\partial\Psi_y}/{\partial\lambda^{x}_i}=0$, {\it i.~e.}, $\Psi\equiv\Psi(S_i,\pi_i,\theta)$.  
This matrix has a zero-mode, which we identify with  the
gauge-symmetry generator. To pull out the hidden symmetry, we force the zero-mode to satisfy the relation
\begin{equation}
\label{matrix02}
\int \,\, d^3y \,\,\nu^{(0)}_\alpha(x)\,\,f_{\alpha\beta}(x,y)= 0,
\end{equation}
which allows us to compute $\Psi$.

Let us start with the symmetry generated by the following zero-mode,
\begin{equation}
\label{3017}
\nu^{(0)}=\begin{pmatrix}
0 \\
S_i \\
0 \\
1
\end{pmatrix}.
\end{equation}

Since this zero-mode and the symplectic matrix~(\ref{3016}) satisfy
Eq.~(\ref{matrix02}), we find that
\begin{equation}
\label{3017a}
\Psi = \frac{1}{2}S_i^2.
\end{equation}

To start the second step we require that the contraction of the
zero-mode with the potential gradient generates no additional
constraints.  The correction terms can be explicitly computed as
functions of $\theta$. Integration yields the correction to first-order in $\theta$:
\begin{equation}
\label{3018}
{\cal G}^{(1)}(S_i,\pi_i,\theta) = -\frac{1}{J} S_i\pi^i\theta.
\end{equation}
Substitution in the last term on the right-hand side of
Eq.~(\ref{3013a}) yields the the new Lagrangian
\begin{eqnarray}
\label{3019}
\tilde {\cal L}^{(0)} = \pi^i\dot{S}_i + \Psi\dot{\theta} - \frac{J}{2}\,(\partial_k S_i)^2 - \frac{1}{2}\,\lambda\,\bigl(S^2_i - 1\bigr)- \frac{1}{2J}\,\pi^2_i +\frac{1}{2J}S_i\pi^i\theta - \sum_{n=2}^\infty{\cal G}^{(n)}.
\end{eqnarray}

At this point the model is not yet gauge invariant because the
contraction of the zero-mode $\nu^{(0)}$ with the gradient of the
potential $V^{(0)}$ is nonzero. This calls for computation of the remaining
correction terms ${\cal G}^{(n)}$ ($n=2,3,\ldots$) as
functions of $\theta$. In practice, to carry out the sum in the last
term on right-hand side of Eq.~\eqref{3019} we only have to demand
that the zero-mode generate no new constraint. The correction term
${\cal G}^{(2)}$ is therefore given by the equality
\begin{equation}
\label{3020}
{\cal G}^{(2)} =  \frac {1}{2J} S_i^2 \theta^2,
\end{equation}
and the correction terms ${\cal G}^{(n)}$ ($n\geq 3$) are
null. Substitution of these results into the first-order
Lagrangian~(\ref{3019}) yields the result
\begin{align}
\label{3021}
{\tilde {\cal L}}^{(0)} =& \pi^i\dot{S}_i + \frac{1}{2}S_i^2\dot{\theta} - \frac{J}{2}\left(\partial_{k}S_i\right)^2 - \frac{1}{2}\,\lambda\,\bigl(S^2_i-1\bigr) - \frac{1}{2J}\pi^2_i
+ \frac{1}{J}S_i\pi^i\theta - \frac{1}{2J}S_i^2\theta^2.
\end{align}

The zero-mode $\nu^{(0)}$ no longer producing new constraints, the
model is symmetric and, in compliance with the symplectic formalism,
the symmetry generator is the zero-mode.

We now want to recover the invariant second-order Lagrangian from the first-order form in Eq.~(\ref{3021}).
To this end, the canonical momenta must be eliminated from the
Lagrangian~(\ref{3021}). The canonical momenta computed from the
equation of motion for the $\pi_i$ are
\begin{equation}
\label{3022}
\pi_i = J\dot{S}_i + S_i\theta.
\end{equation}

When Eq.~\eqref{3022} is inserted in the first-order
Lagrangian~(\ref{3021}), the following expression for the second-order
Lagrangian is obtained:
\begin{equation}
\label{3023}
\tilde {\cal L} = \frac{J}{2}\,\partial_\mu S^i\partial^\mu S_i - (\dot{S_i}S^i)\theta -
\frac{1}{2}\bigl(S^2_i - 1\bigr)\lambda ,
\end{equation}
with the following gauge invariant Hamiltonian,
\begin{equation}
\label{3024}
\tilde {\cal H} = \frac{1}{2J}\pi^2_i + \frac{J}{2}\,(\partial_k S_i)^2
+ \frac{1}{J}\pi^i S_i \theta + \frac{1}{2J} S_i^2 \theta^2 + \frac{\lambda}{2}(S^2_i -1).
\end{equation}

The symplectic formalism identifies the zero-mode with the generator
of the infinitesimal gauge transformations $\delta\tilde
\xi_\alpha^{(0)}=\varepsilon \nu^{(0)}$, namely,
\begin{align}
\label{3025}
\delta S_i &= 0,\nonumber\\
\delta \pi_i &= \varepsilon S_i ,\nonumber\\
\delta \lambda &= 0,\\
\delta \theta &= \varepsilon.\nonumber
\end{align}

These transformations introduce Hamiltonian changes of the form
\begin{equation}
\label{3025a}
\delta{\tilde {\cal H}} = 0.
\end{equation}

Henceforth, we are interested in disclosing the hidden symmetry of the
isotropic antiferromagnet in the original phase space
$(S_i,\pi_i)$. To this end, we will apply the Dirac method to obtain
the set of constraints on the gauge-invariant isotropic
antiferromagnet described by the Lagrangian~(\ref{3023}) and
Hamiltonian~(\ref{3024}). We therefore have that
\begin{align}
\label{3026}
\phi_1 &= \pi_\lambda,\nonumber\\
\phi_2 &= - \frac 12(S^2_i - 1),
\end{align}
and
\begin{align}
\label{3027}
\varphi_1 &= \pi_\theta,\nonumber\\
\varphi_2 &= \frac{1}{J}S_i\pi_i - \frac{1}{J}S_i^2\theta,
\end{align}
where $\pi_\lambda$ and $\pi_\theta$ are the canonical momenta
conjugated to $\lambda$ and $\theta$, respectively. The corresponding
Dirac matrix is singular.  However, a few constraints have
nonvanishing Poisson brackets, which point to both second- and
first-class constraints.  To solve this problem we separate the
former from the latter via constraint
analysis. The set of first-class constraints is
\begin{align}
\label{3028}
\chi_1 &= \pi_\lambda,\nonumber\\
\chi_2 &= - \frac 12 (S^2_i - 1) + \pi_\theta,
\end{align}
while the set of second-class constraints is
\begin{eqnarray}
\label{3029}
\varsigma_1 &=& \pi_\theta,\nonumber\\
\varsigma_2 &=& \frac{1}{J} S_i\pi_i - \frac{1}{J} S^2_i\theta .
\end{eqnarray}

Since the second-class constraints are assumed to be equal to zero in a strong way, the Dirac
brackets are 
\begin{align}
\label{3030}
\lbrace S_i(x), S_j(y)\rbrace^* &= 0,\nonumber\\
\lbrace S_i(x), \pi_j(y)\rbrace^* &= \delta_{ij}\,\delta(x-y),\\
\lbrace \pi_i(x), \pi_j(y)\rbrace^* &= 0.\nonumber
\end{align}

Hence, the gauge invariant Hamiltonian can be rewritten in the form
\begin{eqnarray}
\label{3031}
\tilde {\cal H} &= \dfrac{1}{2J}\pi^2_i + \dfrac{J}{2}\,(\partial_k S_i)^2
- \dfrac{1}{2}\dfrac{(S_i\pi^i)^2}{S_i S^i} + \dfrac{\lambda}{2}(S^2_i -1)\nonumber\\
&= {1\over 2J} \pi_iM^{ij}\pi_j + \dfrac{J}{2}\,(\partial_k S_i)^2 + \dfrac{\lambda}{2}(S_i^2 - 1),
\end{eqnarray}
where the phase space metric $M^{ij}$ is given by the equality
\begin{equation}
\label{3032}
M^{ij} = \delta^{ij} - \dfrac {S^i S^j}{S_k^2},
\end{equation}
which is a singular matrix.  

The set of first class constraints becomes
\begin{align}
  \label{3033}
  \chi_1 &= \pi_\lambda,\nonumber\\
  \chi_2 &= - \frac 12(S^2_i - 1).
\end{align}

The constraint $\chi_2$, originally a second-class constraint, becomes
the generator of gauge symmetries and satisfies the first-class property
\begin{equation}
\label{3035}
\{\chi_2, \tilde{H} \} = 0.
\end{equation}

In view of this result, the infinitesimal gauge transformations are computed as
\begin{align}
  \label{3036}
  \delta S_i &= \varepsilon\lbrace S_i,\chi_2\rbrace = 0,\nonumber\\
  \delta \pi_i &= \varepsilon\lbrace\pi_i,\chi_2\rbrace=\varepsilon S_i,\\
  \delta\lambda &= 0,\nonumber
\end{align}
where $\varepsilon$ is an infinitesimal. 

It is easy to verify that the Hamiltonian~(\ref{3031}) is invariant
under these transformations because the $S_i$ are eigenvectors of the
phase-space metric $M_{ij}$ with null eigenvalues.

\section{The functional Schr\"odinger equation}
\label{s4}

The $2D$ isotropic antiferromagnet is described by the $O(3)$ nonlinear sigma model given by the Lagrangian
\begin{align}
L &= \frac{J}{2} \int{d^{2}x \partial_\mu S^a \partial^\mu S^a} \nonumber \\
&= \frac{J}{2} \int{d^{2}x [(\partial_0 S^a)^2 - (\partial_i S^a)^2]
}\qquad(\mu = 0,1,2;\, i = 1,2),
\label{H1}
\end{align}
with the constraint
\begin{equation}
\label{H2}
\Omega_1 = S^a S^a - 1.
\end{equation}
Here $a$ is an index related to the $O(3)$ symmetry group, and the metric has signature $(+,-,-)$.

To start our search for the functional Schr\"odinger we
strongly impose the constraint $\Omega_1$ in Eq.~(\ref{H2}). This lets
us write one of the fields, say $S^3$, in terms of the fields
$S^1$ and $S^2$:
\begin{equation}
\label{H3}
S^3 = \sqrt{1 - S^i S^i}\qquad(i=1, 2).
\end{equation}

From Eq.~(\ref{H3}), we obtain the result
\begin{equation}
\label{H4}
\partial_\mu S^3 = - \frac{S^i \partial_\mu S^i}{\sqrt{1 - S^i S^i}}.
\end{equation}

Introducing Eqs.~(\ref{H3}) and~(\ref{H4}) in Eq.~(\ref{H1}), we
express the model in the fields $S^1$ and $S^2$:
\begin{equation}
L = \frac{J}{2} \int{d^{2}x g_{ij} \partial_\mu S^i \partial^\mu S^j}, 
\label{H5}
\end{equation}
where
\begin{equation}
g_{ij} = \delta_{ij} + \frac{S_i S_j}{1 - S^i S^i}. 
\label{H6}
\end{equation}

Let us now construct the model Hamiltonian, for subsequent quantization. We compute the momenta
\begin{equation}
\pi_i = \frac{\partial L}{\partial(\dot{S^i})}. 
\label{H7}
\end{equation}

Then, we have that
\begin{equation}
\pi_i = J g_{ij}\dot{S^j}. 
\label{H8}
\end{equation}

In order to write the model in its Hamiltonian form, we must invert
Eq.~(\ref{H8}), so that we can express the {\it velocities} in terms
of the momenta. The computation of the inverse of $g_{ij}$ gives us

\begin{equation}
\tilde{g}^{ij} = \delta^{ij} - S^i S^j,
\label{H9}
\end{equation}
so that
\begin{equation}
\dot{S^i} = \frac{1}{J}\tilde{g}^{ij} \pi_j.
\label{H10}
\end{equation}

The model Hamiltonian, the general expression for which is
\begin{equation}
H = \int{d^{2}x (\pi_i \dot{S}^i - L)},
\label{H11}
\end{equation}
takes the particular form,
\begin{equation}
H = \int{d^{2}x \left(\frac{1}{2J}\tilde{g}^{ij} \pi_i \pi_j + \frac{J}{2} g_{ij} \partial_k S^i \partial_k S^j\right)},
\label{H12}
\end{equation}
where $\partial_k$ denotes the partial derivatives with respect to the
spatial coordinates. By definition, the $(S^i, \pi_j)$ form
canonically conjugated pairs, with the usual Poisson brackets,
\begin{equation}
\{S_i (x), \pi_j (y)\} = \delta_{ij} \delta^2 (x-y).
\label{H13}
\end{equation}

To derive functional Schr\"odinger equation for the $2D$ isotropic
antiferromagnet we introduce the wave-functional
$\Psi[S^i, t]$ and treat $S^i$ and $\pi_i$ as quantum
operators. In other words, in the field representation the momenta
are replaced by the following functional derivatives:
\begin{equation}
\label{H14}
\pi_i (x) \longrightarrow -i\frac{\delta}{\delta S^i(x)},
\end{equation}
where we have set $\hbar = 1$.

The wave-functional $\Psi$ satisfies the functional Schr\"odinger equation
\begin{equation}
\label{H15}
i\frac{\partial}{\partial t}\Psi[S^i, t] = \hat{H}[S^i,t]\Psi[S^i, t]  ,
\end{equation}
where $ \hat{H}$ is the operator version of the Hamiltonian in Eq.~(\ref{H12}).

Since $\tilde{g}^{ij}$ depends on the fields, the kinetic term in the
on the right-hand side of Eq.~(\ref{H12}) will give rise to factor-ordering
ambiguities upon quantization. To resolve these ambiguities, we
choose a particular factor-ordering: we write all field
functions to the left of the momenta operators. To justify this choice
we note that in the study of the first-class constrained version of the
model the ordering is consistent with the operator version of the
classical-constraint algebra. Moreover, the first-class Hamiltonian
that will be derived below leads to the same functional Schr\"odinger
equation.

Given Eq.~(\ref{H12}) and the aforementioned particular choice of
factor ordering, we obtain the following functional Schr\"odinger
equation for the isotropic antiferromagnet:
\begin{equation}
\int{d^{2}x \left(\frac{1}{2J}\tilde{g}^{ij} \frac{\delta^2
      \Psi}{\delta S^i \delta S^j} + \frac{J}{2} g_{ij} \partial_k
    S^i \partial_k S^j \Psi\right)} 
= i\frac{\partial}{\partial t} \Psi.
\label{H16}
\end{equation}

Since the Hamiltonian~(\ref{H12}) does not explicity depend on time,
we may factor out the time dependence of the wave-functional and
write the equality
\begin{equation}
\Psi[S^i, t] = e^{-Et} \Psi[S^i].
\label{H16a}
\end{equation}

In view of Eq.~(\ref{H16}) we then see that $\Psi[S^i]$ must satisfy
the time-independent functional Schr\"odinger equation
\begin{equation}
\int{d^{2}x (\frac{1}{2J}\tilde{g}^{ij} \frac{\delta^2 \Psi}{\delta S^i \delta S^j} + \frac{J}{2} g_{ij} \partial_k S^i \partial_k S^j \Psi)} = E \Psi.
\label{H16b}
\end{equation}

It is clear that the solution of Eq.~(\ref{H16b}) will yield the
energies $E$ for the studied model.

Let us now consider the $2D$ isotropic aniferromagnet as a
first-class constrained field theory. We also want to write the
functional Schr\"odinger equation for our model. To this end, we
use Dirac's prescription to canonically quantize first-class
constrained systems \cite{Dirac} . As we shall see below, the
functional Schr\"odinger equation will be the same as the one in
Eq.~(\ref{H16}). It proves convenient to write the first-class Hamiltonian of the
model in the following way:
\begin{equation}
\label{H17}
H = \int{d^2x \left[\frac{1}{2J}\bar{g}^{ab}\pi_a \pi_b + \frac{J}{2} \partial_i S^a \partial_i S^a - \lambda(S^2_a - 1) + v_\lambda \pi_\lambda\right]},
\end{equation}
where 
\begin{equation}
\bar{g}^{ab} = \delta^{ab} - \frac{S^a S^b}{S^a S^a}. 
\label{H18}
\end{equation}

We note that $\pi_\lambda=0$ is a first-class constraint and
$v_\lambda$ is a Lagrangian multiplier. The formulation
via Eq.~(\ref{H17}) is classically equivalent to the initial one, via
Eq.~(\ref{H12}). In the appropriate gauge, the equations of motion for the
physical fields in Eq.~(\ref{H17}) and in Eq.~(\ref{H12}) are the same
\cite{Gil}.

Only one modification of the functional Schr\"odinger method we have
described is necessary to comply with Dirac's prescription for
first-class constrained systems. The wave-functional must be
annihilated by the operator version of the constraints, besides
satisfying the functional Schr\"odinger equation \cite{Jackiw, Gil2}.
In our case, the requirement that the operator version of the
constraint $\chi_2$, Eq.~(\ref{3033}), annihilate the wave-functional
imposes no condition upon $\Psi$. We therefore make the canonical
transformation \cite{Gil},
\begin{equation}
\label{f4}
S^3 \longrightarrow - \pi_3,\quad\pi_3 \longrightarrow S^3,
\end{equation}
which changes $\chi_2$ and $H$ to
\begin{align}
\label{f5}
\tilde{\chi_2} &= \pi_3 \pi_3 + S^i S^i - 1 = 0, \nonumber \\
\tilde{H} = \int d^2x\, \Big\{&\frac{1}{2J}[\pi_i \pi_i - (\frac{S^i S^j}{\pi_3 \pi_3 + S^k S^k}) \pi_i \pi_j ] + \frac{J}{2} [\partial_x S^i \partial_x S^i + \partial_x \pi_3 \partial_x \pi_3] \nonumber \\
&+ \frac{1}{2J} [S^3 S^3 + 2(\frac{S^i S^3}{\pi_3 \pi_3 + S^k S^k})\pi_i \pi_3 - (\frac{S^3 S^3}{\pi_3 \pi_3 + S^k S^k})\pi_3 \pi_3] \nonumber \\
&- \lambda (\pi_3 \pi_3 + S^i S^i - 1) + v_\lambda \pi_\lambda \Big\}.
\end{align}

We are now ready to write the equations for the wave-functional
$\Psi[S^3,S^i,\lambda]$. The first two will be obtained from the
operator version of the constraints $\pi_\lambda = 0$,
Eq.~(\ref{3033}), and $\tilde{\chi_2}$, Eq.~(\ref{f5}), which
annihilate $\Psi$. The last one is the functional Schr\"odinger
equation, which will be derived from the operatorial version of the
Hamiltonian ($\hat{\tilde{H}}$), Eq.~(\ref{f5}). Thus, in the field
representations we have that
\begin{align}
\label{f6}
\frac{\delta \Psi}{\delta \lambda} = 0&, \\
\label{f7}
-\frac{\delta^2 \Psi}{\delta (S^3)^2} + (S^i S^i - 1)\Psi &= 0, \\
i \frac{\partial \Psi}{\partial t} = \int d^2x \Big[&\frac{1}{2J}(-\frac{\delta^2 \Psi}{\delta (S^i)^2} + (S^i S^j) \frac{\delta^2 \Psi}{\delta S^i \delta S^j}) \nonumber \\
\label{f8}
&+ \frac{J}{2} (\partial_x S^i \partial_x S^i \Psi - \partial_x \frac{\delta}{\delta S^3}\partial_x \frac{\delta}{\delta S^3} \Psi) \nonumber \\
&+ \frac{1}{2J}( S^3 S^3 \Psi - 2(S^i S^3) \frac{\delta^2 \Psi}{\delta S^i \delta S^3} + (S^3 S^3)\frac{\delta^2 \Psi}{\delta (S^3)^2}\Big],
\end{align} 
where we have explicity used Eq.~(\ref{f7}) to deal with the operators
in the denominators of the fractions in $\hat{\tilde{H}}$. The
particular choice of factor ordering in Eqs.~(\ref{f6}) -~(\ref{f8})
preserves the classical-constraint algebra and the involution of the
constraints with the Hamiltonian.

From Eqs.~(\ref{f6}) and~(\ref{f7}) we have that
\begin{align}
\label{f9}
\Psi[S^i, S^3, \lambda, t] &= \Psi[S^i, S^3, t], \\
\label{f10}
\Psi[S^i, S^3, t] &= \exp [ \int {d^2y S^3 \sqrt{S^i S^i - 1}}] \Psi_{phys}[S^i, t]. 
\end{align}

Substitution of the right-hand side of Eq.~(\ref{f10}) for $\Psi$ in
the functional Schr\"odinger equation~(\ref{f8}) yields the equality
\begin{equation}
\int{d^{2}x (\frac{1}{2J}\tilde{g}^{ij} \frac{\delta^2 \Psi_{phys}}{\delta S^i \delta S^j} + \frac{J}{2} g_{ij} \partial_k S^i \partial_k S^j \Psi_{phys})} = E \Psi_{phys},
\label{f11}
\end{equation}
where $g_{ij}$ and $\tilde{g}^{ij}$ are given by Eqs.~(\ref{H6})
and~(\ref{H9}), respectively, and the time dependence of $\Psi_{phys}$
is given by Eq.~(\ref{H16a}). A few
terms proportional to the Dirac delta function at the point
zero, \emph{i.~e.}, $\delta(0)$, appear in the derivation of the
Eq.~(\ref{f11}). These terms contribute energy infinities,
which can be removed by the usual regularization techniques.

Comparison of Eq.~(\ref{f11}) with Eq.~(\ref{H16b}) shows that they
are identical.

\section{Conclusions}\label{s5}

In this work we have obtained the Dirac brackets of the
two-dimensional isotropic antiferromagnet by two different ways: via
the usual Dirac formalism and the symplectic formalism. We have
used the symplectic embedding formalism to unveil
hidden symmetries in the 2D isotropic antiferromagnet. We
have obtained the gauge-invariant Lagrangian and Hamiltonian for the
system. 

In this context, the equivalence between the $O(3)$ non-linear sigma
model and the $CP^1$ model deserves mention \cite{Banerjee}. In fact,
the dynamical variables of the $CP^1$ model are the pair of complex
fields $Z(x) = (Z_1(x), Z_2(x))$ which are constrained to lie on the
unit three-sphere $S^3$: $Z^* Z = |Z|^2 = |Z_1|^2 + |Z_2|^2 = 1$. This
model is described by the non-linear Lagrangian ${\cal L}
= \partial_\mu Z^* \partial^\mu Z - (Z^* \partial_\mu
Z)(Z \partial^\mu Z^*)$ and it is invariant under a local $U(1)$ gauge
symmetry. Using the Hopf bundle \cite{Hopf}, $S^a = Z^* \sigma^a Z$,
which characterizes maps from $S^3$ to $S^2$, where the $\sigma^a$ are
the Pauli matrices, we can see that the Lagrangian~(\ref{3001}) is
equivalent to the Lagrangian of the $CP^1$ model. In view of this, the
existence of a hidden gauge symmetry in the model could have been
expected.

For a particular choice of factor ordering, we wrote the
functional Schr\"odinger equation for the first-class Hamiltonian and
for the original Hamiltonian and showed that they are all identical,
which justifies our factor-ordering choice. In a future paper we
intend to present the gauge-invariant version for the $2D$ anisotropic
antiferromagnet, as well as its functional Schr\"odinger equation,
and discuss the spectra of both the isotropic and anisotropic models.

\section{ Acknowledgments}
The authors would like to thank the CNPq and CAPES for financial support.

\appendix

\section{Symplectic formalism}
\label{A1}
In this appendix, we will present a brief review of the symplectic
formalism. Let us consider a general first-order Lagrangian described
by the symplectic variables $\xi^i$ and their generalized canonical
momenta $a_i$,
\begin{equation}
\label{a1}
{\cal L} = a_{i}(\xi)\dot{\xi} - {\cal H}(\xi),
\end{equation}
which is obtained from a conventional second-order Lagrangian by
introducing certain auxiliary fields and performing the Legendre
transformation.  The symplectic potential $V(\xi)$ is none other than
the Hamiltonian ${\cal H}(\xi)$, as we can see by Legendre
transforming the first-order Lagrangian in Eq.~(\ref{a1}).  Thus, the
Lagrangian in Eq.~(\ref{a1}) may be rewritten in the form
\begin{equation}
\label{a2}
{\cal L}dt = a_{i}(\xi)d\xi - {\cal H}(\xi)dt,
\end{equation}
and the first term on the right side defines the canonical one-form $a_{i} d\xi^i \equiv a(\xi)$. Using the variational principle, we obtain the dynamical equations of motion
\begin{equation}
\label{a3}
f_{ij}\dot{\xi}^j = \frac{\partial}{\partial \xi^i} {\cal H}(\xi),
\end{equation}
where
\begin{equation}
\label{a4}
f_{ij} = \frac{\partial a_j}{\partial \xi^i} - \frac{\partial a_i}{\partial \xi^j},
\end{equation}
which is called the symplectic two-form, $\frac{1}{2} f_{ij} d\xi^i
d\xi^j \equiv f(\xi)$. 

Usually, the geometric structure of the theory
is fully determined by the generalized canonical momenta $a_i(\xi)$
and is insensitive to the functional form of $V(\xi)$. The symplectic
matrix $f_{ij}$ gives the geometric structure in phase
space. Theories are classified as unconstrained or constrained,
depending on whether $f_{ij}$ has an inverse or not, respectively.

In the unsconstrained case, we can directly obtain equations of
motion such that 
\begin{equation}
\label{a5}
\dot{\xi}^i = f^{ij} \frac{\partial {\cal H}(\xi)}{\partial \xi^j}.
\end{equation}
In this case, we can obtain the generalized symplectic brackets as
\begin{equation}
\label{a6}
\dot{\xi}^i =  \lbrace \xi^i,{\cal H}(\xi)\rbrace = \lbrace \xi^i,\xi^j\rbrace \frac{\partial {\cal H}(\xi)}{\partial \xi^j}.
\end{equation}
Comparing Eq.~(\ref{a5}) with Eq.~(\ref{a6}), we obtain the relations between the symplectic two-form matrix and the generalized symplectic bracket

\begin{equation}
\label{a7}
f^{ij} =  \lbrace \xi^i,\xi^j\rbrace ,
\end{equation}
which correspond to the Dirac brackets.

In the more interesting case, when $f_{ij}$ is singular and
constraints arise such that
\begin{equation}
\label{a8}
\Omega^{(a)} = \tilde{\nu}^{(a)i} \frac{\partial}{\partial \xi^i}{\cal H}(\xi) = 0,
\end{equation}
where $\tilde{\nu}^{(a)i}$ are called zero-modes, the superscript $i$
corresponds to the symplectic variables $\xi^i$, and $(a)$ denotes the
number of constraints. Furthermore, from the persistence in time of
the constraints in Eq.~(\ref{a8}), we have that
\begin{equation}
\label{a9}
\dot{\Omega}^{(a)} = \frac{\partial\Omega^{(a)}}{\partial \xi^i}\dot{\xi}^i = 0.
\end{equation}

When this happens, we can modify part of the canonical sector to make
the symplectic two-form matrix invertible. To this end, we introduce
appropriate Lagrange multipliers and incorporate the requirement in
Eq.~(\ref{a9}), that constraints in the first-order Lagrangian must be
stable under time evolution. Thus, the modified first-order
Lagrangian is described by the expression
\begin{equation}
\label{a10}
{\cal L}^{(k)} = a_{i}^{(k)}(\xi) \dot{\xi}^{(k)i} + \Omega_{i}^{(k)}\dot{\alpha}^{(k)i} - {\cal H}^{(k)}(\xi),
\end{equation}
where the integer $k$ in the superscript denotes the iteration number
to generate the modified nonsigular symplectic matrix, and the
$k$-th iteration Hamiltonian, 
\begin{equation}
\label{a11}
{\cal H}^{(k)}(\xi) = {\cal H}^{(k-1)}(\xi)\Big|_{\Omega_{i}^{(k-1)}=0},
\end{equation}
corresponds to the reduced Hamiltonian of the theory.

If we come to a nonsingular $f_{ij}$ after a finite number of
iterations, we interrupt the iterative sequence and obtain the Dirac
brackets from the inverse of the matrix $f_{ij}$.  Otherwise, the
sequence grows to infinity, in which case the zero-mode plays an
important role, generating a gauge symmetry, and the transformation
rules are given by the zero-mode, such that
\begin{equation}
\label{a12}
\delta \xi^{(k)i} = \varepsilon \tilde{\nu}^{(k)i} ,
\end{equation} 
where $\varepsilon$ is a function of time. 

In this step of the method, we then need some gauge-fixing conditions,
which are a kind of constraint. These constraints must be introduced
in the canonical sector of the first-order Lagrangian in
Eq.~(\ref{a10}). Following the prescription of the symplectic
formalism as already described, we obtain the Dirac brackets, which
are the bridge to the quantum commutators.

\section{Brief review of the general theory of symplectic 
embedding}
\label{A2}

This appendix closely follows the ideas in Ref.~\cite{Oliveira}.

Consider a general noninvariant mechanical model whose dynamics is
governed by a Lagrangian ${\cal L}(a_i,\dot a_i,t)$,
($i=1,2,\dots,N$), where $a_i$ and $\dot a_i$ are the space and
velocities variables, respectively. Notice that this model results in
no loss of generality or physical content. In the symplectic
method the zeroth-iterative first-order Lagrangian one-form is written
as
 \begin{equation}
\label{2000}
{\cal L}^{(0)}dt = A^{(0)}_\theta d\xi^{(0)\theta} - V^{(0)}(\xi)dt,
\end{equation}
and the symplectic variables are
\be
\xi^{(0)\theta} =  \left\{ \begin{array}{ll}
                               a_i, & \mbox{with $\theta=1,2,\dots,N $} \\
                               p_i, & \mbox{with $\theta=N + 1,N + 2,\dots,2N ,$}
                           \end{array}
                     \right.
\ee
where $A^{(0)}_\theta$ are the canonical momenta and $V^{(0)}$ is the
symplectic potential. From the Euler-Lagrange equations of motion we
obtain the symplectic tensor
\begin{align}
\label{2010}
f^{(0)}_{\theta\beta} = {\partial A^{(0)}_\beta\over \partial \xi^{(0)\theta}}
-{\partial A^{(0)}_\theta\over \partial \xi^{(0)\beta}}.
\end{align}

If the two-form $f \equiv \frac{1}{2}f_{\theta\beta}d\xi^\theta \wedge
d\xi^\beta$ is singular, the symplectic matrix~(\ref{2010}) has a
zero-mode $\nu^{(0)}$ that generates a new constraint when it is
contracted with the gradient of the symplectic potential,
\begin{equation}
\label{2020}
\Omega^{(0)} = \nu^{(0)\theta}\frac{\partial V^{(0)}}{\partial\xi^{(0)\theta}}.
\end{equation}

We introduce this constraint in the zero-iteration Lagrangian
one-form~(\ref{2000}) via a Lagrange multiplier $\eta$ to generate the next one
\begin{align}
\label{2030}
{\cal L}^{(1)}dt &= A^{(0)}_\theta d\xi^{(0)\theta} + d\eta\Omega^{(0)}- V^{(0)}(\xi)dt,\nonumber\\
&= A^{(1)}_\gamma d\xi^{(1)\gamma} - V^{(1)}(\xi)dt\qquad(\gamma=1,2,\dots,2N + 1),\end{align}
and
\begin{align}
\label{2040}
V^{(1)}&=V^{(0)}|_{\Omega^{(0)}= 0},\nonumber\\
\xi^{(1)_\gamma} &= (\xi^{(0)\theta},\eta),\\
A^{(1)}_\gamma &=(A^{(0)}_\theta, \Omega^{(0)}).\nonumber
\end{align}

The first-iteration symplectic tensor is then
\begin{align}
\label{2050}
f^{(1)}_{\gamma\beta} = {\partial A^{(1)}_\beta\over \partial \xi^{(1)\gamma}}
-{\partial A^{(1)}_\gamma\over \partial \xi^{(1)\beta}}.
\end{align}

If this tensor is nonsingular, we stop the iterative process and
obtain the Dirac brackets among the phase space variables from the
inverse matrix $(f^{(1)}_{\gamma\beta})^{-1}$. Consequently, the
Hamilton equation of motion can be computed and solved, as discussed
in Ref.~\cite{gotay}. It is well known that a physical system can be
described at least classically in terms of a symplectic manifold
$M$. From the physical point of view, $M$ is the phase space of the
system, while a nondegenerate closed 2-form $f$ can be identified with
the Poisson bracket. To fix the dynamics of the system we only have to
specify a real-valued function (Hamiltonian) $H$ in phase space. In
other words, one of these real-valued function solves the Hamilton
equation, namely, 
\begin{equation}
\label{2050a1} \iota(X)f=dH,
\end{equation}
and determines the classical dynamical trajectories of the system in phase space.

If $f$ is nondegenerate, Eq.~(\ref{2050a1}) has an unique
solution. The nondegeneracy of $f$ means that the linear map
$\flat:TM\rightarrow T^*M$, defined by $\flat(X):=\flat(X)f$, is an
isomorphism. Equation~(\ref{2050a1}) therefore has a unique solution
for any Hamiltonian $[X=\flat^{-1}(dH)]$. If, by contrast, the tensor
has a zero-mode, a new constraint arises, and the
iterative process goes on until the symplectic matrix become
nonsingular or singular. If this matrix is nonsingular, we will be
able to determine the Dirac brackets. Reference~\cite{gotay} considers the
case of degenerate $f$ in detail. 

The central idea in this embedding formalism is to introduce extra
fields into the model in order to obstruct the solutions of the
Hamiltonian equations of motion.  We introduce two arbitrary functions
that depend on the original phase space and on WZ variables,
namely, $\Psi(a_i,p_i)$ and $G(a_i,p_i,\eta)$, in the first-order
Lagrangian one-form as follows:
\begin{equation}
\label{2060a}
{\tilde{\cal L}}^{(0)}dt = A^{(0)}_\theta d\xi^{(0)\theta} + \Psi d\eta - {\tilde V}^{(0)}(\xi)dt,
\end{equation}
with
\begin{equation}
\label{2060b}
{\tilde V}^{(0)} = V^{(0)} + G(a_i,p_i,\eta),
\end{equation}
where the arbitrary function $G(a_i,p_i,\eta)$ is expanded in the WZ field, given by
\begin{equation}
\label{2060}
G(a_i,p_i,\eta)=\sum_{n=1}^\infty{\cal G}^{(n)}(a_i,p_i,\eta),\qquad{\cal G}^{(n)}(a_i,p_i,\eta)\sim\eta^n\,,
\end{equation}
and satisfies the following boundary condition:
\begin{align}
\label{2070}
G(a_i,p_i,\eta=0) = 0.
\end{align}

We extend the symplectic variables to include the WZ variable $\tilde\xi^{(0)\tilde\theta} = (\xi^{(0)\theta},\eta) ( {\tilde\theta}=1,2,\dots,2N+1$) and the first-iterative symplectic potential becomes
\begin{equation}
\label{2075}
{\tilde V}^{(0)}(a_i,p_i,\eta) = V^{(0)}(a_i,p_i) + \sum_{n=1}^\infty{\cal G}^{(n)}(a_i,p_i,\eta).
\end{equation}

In this context, the new canonical momenta are
\begin{equation}
  {\tilde A}_{\tilde\theta}^{(0)} =
  \begin{cases}
    A_{\theta}^{(0)}, & (\tilde\theta =1,2,\dots,2N)\\
    \Psi, & (\tilde\theta= 2N + 1)
  \end{cases}
\end{equation}
and the new symplectic tensor is
\begin{equation}
{\tilde f}_{\tilde\theta\tilde\beta}^{(0)} = \frac {\partial {\tilde A}_{\tilde\beta}^{(0)}}{\partial \tilde\xi^{(0)\tilde\theta}} - \frac {\partial {\tilde A}_{\tilde\theta}^{(0)}}{\partial \tilde\xi^{(0)\tilde\beta}},
\end{equation}
that is,
\begin{equation}
\label{2076b}
{\tilde f}_{\tilde\theta\tilde\beta}^{(0)} = 
\begin{pmatrix}
 { f}_{\theta\beta}^{(0)} & { f}_{\theta\eta}^{(0)}
\cr { f}_{\eta\beta}^{(0)} & 0
\end{pmatrix}.
\end{equation}

To sum up, we have two steps: in the first we compute $\Psi(a_i,p_i)$,
while in the second we calculate $G(a_i,p_i,\eta)$. At the beginning 
the first step we impose that this new symplectic tensor
(${\tilde f}^{(0)}$) have a zero-mode $\tilde\nu$, which leads
to the following condition:
\begin{equation}
\label{2076}
\tilde\nu^{(0)\tilde\theta}{\tilde f}^{(0)}_{\tilde\theta\tilde\beta} = 0.
\end{equation}

At this point, $f$ becomes degenerate. Consequently, we introduce an
obstruction to solve, in an unique way, the Hamilton equation of
motion in Eq.~(\ref{2050a1}). Assuming that the zero-mode
$\tilde\nu^{(0)\tilde\theta}$ is
\begin{equation}
\label{2076a}
\tilde\nu^{(0)}=
\begin{pmatrix}
\mu^\theta & 1
\end{pmatrix},
\end{equation}
and using the relation in~(\ref{2076}) together with~(\ref{2076b}), we
find a set of equations, namely, 
\begin{equation}
\label{2076c}
\mu^\theta{ f}_{\theta\beta}^{(0)} + { f}_{\eta\beta}^{(0)} = 0,
\end{equation}
where
\begin{equation}
{ f}_{\eta\beta}^{(0)} =  \frac {\partial A_\beta^{(0)}}{\partial \eta} - \frac {\partial \Psi}{\partial \xi^{(0)\beta}}.
\end{equation}

The matrix elements $\mu^\theta$ are chosen to disclose the desired
gauge symmetry. In this formalism the zero-mode
$\tilde\nu^{(0)\tilde\theta}$ is the gauge-symmetry generator. This
characteristic is important because it opens the possibility of
disclosing the desired hidden gauge symmetry from the noninvariant
model. It enables the symplectic embedding formalism to
deal with noninvariant systems. From Eq.~(\ref{2076})
we obtain differential equations involving $\Psi(a_i,p_i)$,
Eq.~(\ref{2076c}), and after straightforward computation,
determine $\Psi(a_i,p_i)$.

To compute $G(a_i,p_i,\eta)$ in the second step, we impose that no
additional constraints arise from the contraction of the zero-mode
$(\tilde\nu^{(0)\tilde\theta})$ with the gradient of the potential
${\tilde V}^{(0)}(a_i,p_i,\eta)$. This condition generates a general
differential equation, which reads
\begin{widetext}
\begin{align}
\label{2080}
\tilde\nu^{(0)\tilde\theta}\frac{\partial {\tilde
    V}^{(0)}(a_i,p_i,\eta)}{\partial{\tilde\xi}^{(0)\tilde\theta}}\,&=
\, 0,\\
\mu^\theta \frac{\partial {V}^{(0)}(a_i,p_i)}{\partial{\xi}^{(0)\theta}} + \mu^\theta \frac{\partial {\cal G}^{(1)}(a_i,p_i,\eta)}{\partial{\xi}^{(0)\theta}} 
\,+\, \mu^\theta\frac{\partial {\cal G}^{(2)}(a_i,p_i,\eta)}{\partial{\xi}^{(0)\theta}} +\ldots& \nonumber\\
\,+\,\frac{\partial {\cal G}^{(1)}(a_i,p_i,\eta)}{\partial\eta} + \frac{\partial {\cal G}^{(2)}(a_i,p_i,\eta)}{\partial\eta} + \dots &= 0.
\end{align}
\end{widetext}

Equations~\eqref{2080}~and \eqref{2090}
allow us to compute all correction terms ${\cal
  G}^{(n)}(a_i,p_i,\eta)$ in order of $\eta$. Since this polynomial
expansion in $\eta$ is equal to zero, the coefficient of each power of
$\eta$ must vanish identically. This determines each correction term
of order $\eta^{n}$. For the linear correction term, we
have that
\begin{equation}
\label{2090}
\mu^\theta\frac{\partial V^{(0)}(a_i,p_i)}{\partial\xi^{(0)\theta}} + \frac{\partial{\cal
 G}^{(1)}(a_i,p_i,\eta)}{\partial\eta} = 0.
\end{equation}

For the quadratic term, we find that
\begin{equation}
\label{2095}
{\mu}^{\theta}\frac{\partial{\cal G}^{(1)}(a_i,p_i,\eta)}{\partial{\xi}^{(0)\theta}} + \frac{\partial{\cal G}^{(2)}(a_i,p_i,\eta)}{\partial\eta} = 0.
\end{equation}

More generally, the following recursive equation for $n\geq 2$ results:
\begin{equation}
\label{2100}
{\mu}^{\theta}\frac{\partial {\cal G}^{(n - 1)}(a_i,p_i,\eta)}{\partial{\xi}^{(0)\theta}} + \frac{\partial{\cal
 G}^{(n)}(a_i,p_i,\eta)}{\partial\eta} = 0,
\end{equation}
which allows us to compute the remaining correction terms in order of
$\eta$. 

This process is iterated until~(\ref{2080}) vanish identically so that
the extra term $G(a_i,p_i,\eta)$ can be explicitly obtained. The
gauge-invariant Hamiltonian, identified with the symplectic potential,
is obtained in the form
\begin{equation}
\label{2110}
{\tilde{\cal  H}}(a_i,p_i,\eta) = V^{(0)}(a_i,p_i) + G(a_i,p_i,\eta),
\end{equation}
and the zero-mode ${\tilde\nu}^{(0)\tilde\theta}$ is identified with the generator of an infinitesimal gauge transformation, given by
\begin{equation}
\label{2120}
\delta{\tilde\xi}^{\tilde\theta} = \varepsilon{\tilde\nu}^{(0)\tilde\theta},
\end{equation}
where $\varepsilon$ is an infinitesimal.

\end{document}